# Realizing a Four-Step Molecular Switch in Scanning Tunneling Microscope Manipulation of Single Chlorophyll-a Molecules

**Violeta Iancu, &  Saw-Wai Hla***

*Nanoscale & Quantum Phenomena Institute, Department of Physics and Astronomy, Ohio University, Athens, Ohio, 45701, USA*

----------------------------------------------------------------

**Single chlorophyll-a molecules, a vital resource for the sustenance of life on Earth, have been investigated by using scanning-tunneling-microscope manipulation and spectroscopy on a gold substrate at 4.6 K. The chlorophyll-a binds on Au(111) via its porphyrin unit while the phytyl-chain is elevated from the surface by a support of four $CH_3$ groups. By injecting tunneling electrons from the STM-tip, we are able to bend the phytyl-chain, which enable switching of four molecular conformations in a controlled manner. Statistical analyses and structural calculations reveal that all reversible switching mechanisms are initiated by a single tunnelling-electron energy-transfer process, which induces bond rotation within the phytyl-chain.**

* Corresponding author : Email: hla@ohio.edu
   http://www.phy.ohiou.edu/~hla

Since beginning of 1990, scanning tunneling microscope (STM) has been demonstrated to be a useful tool to manipulate single atoms/molecules on supporting substrates with atomic scale precision (1-4). To date, most STM manipulation experiments are performed on single atoms or small molecules, and the applications of this technique is concentrated mainly in the materials science research area. In this report, we extend the STM manipulation procedures to biological related research area and investigate a relatively large plant molecule known as chlorophyll-a.

Chlorophyll-a induces green color in plant leaves and is a key ingredient in photosynthesis; one of the most important biological processes that converts sunlight into chemical energy in plants (5-9). Chlorophyll-a is also important from the evolutionary standpoint because photosynthesis plays a central role in the early development of life on Earth (6). But, just as chlorophyll has been vital in the development and sustenance of plants and life forms, it may prove to be just as essential to the advancement of "green" energy research and nanotechnology. Because of their non-toxic nature and their abundance in the natural world, plant molecules like chlorophyll-a are given special interest in the quest for "green" energy resources and development of environmental friendly nanoscale devices (10-12). Chlorophyll-a consists of two main components: a porphyrin unit as the 'head' and a long carbon-chain as the 'tail'. In the light harvesting reaction centers found in plant leaves, chlorophyll-a conforms into various shapes by bending the phytyl tail. Molecular conformation is a key process in many biological functions, and controlling conformational changes of biological molecules with a sub-molecular precision is a dream for many scientists. Here, we are not only able to resolve the structure of single chlorophyll-a molecules for the first time, but also able to reversibly switch four molecular conformation in a controlled manner, and the detailed switching mechanisms are explained by means of both experimental analyses and theoretical calculations.

Our experiments were performed by using a home-built low-temperature STM operated at 4.6 K in an ultra-high-vacuum environment (2). Au(111) surface was chosen as a supporting substrate for the experiments because of its relative inertness. The sample surface was cleaned by repeated cycles of sputtering with neon ions and annealing. After confirming cleanliness of the Au (111) sample by STM imaging, a sub-monolayer coverage of chlorophyll-a produced from spinach (Sigma Aldrich, 97% purity) was deposited onto the surface held at room temperature via vacuum evaporation (13). The sample temperature was then lowered to 4.6 K for the measurements. The STM-tip, an electrochemically etched tungsten wire, was prepared by using a controlled tip-crash procedure described in the reference (14) prior to the experiment.

Chlorophyll-a is weakly bound to the Au(111) surface and the molecules are easily displaced during imaging. The STM images of chlorophyll-a on Au(111) show both single molecules and regions of self-assembled molecular clusters preferentially located at the elbows of Au(111) herringbone reconstruction (Fig. 1*a*). To understand the adsorption geometry of chlorophyll-a, we will first discuss the molecule structure in the clusters. The chlorophyll-a clusters grow epitaxially on Au(111) and form a close-packed structure with a unit cell length of 1.6 nm (Fig. 1*a*). Inside the clusters, the molecules position in pairs with their 'heads' facing each other. In a single row along the long-molecular axis direction, the molecules assemble in an alternating 'head-tail-tail-head' arrangement (Fig. 1*b*). From the atomically resolved STM images, the orientation of the molecules with respect to the Au(111) surface is determined. The long molecular axis is aligned along a direction 7±1° deviated from the surface close-packed atomic rows, i.e. [110] surface directions of Au(111). Such close-packed self-assembly of chlorophyll-a is significant, which mimics the in-vivo packing of chlorophyll-a in the photosynthetic membrane and can have important applications in solar-cell and medical devices (10,12).

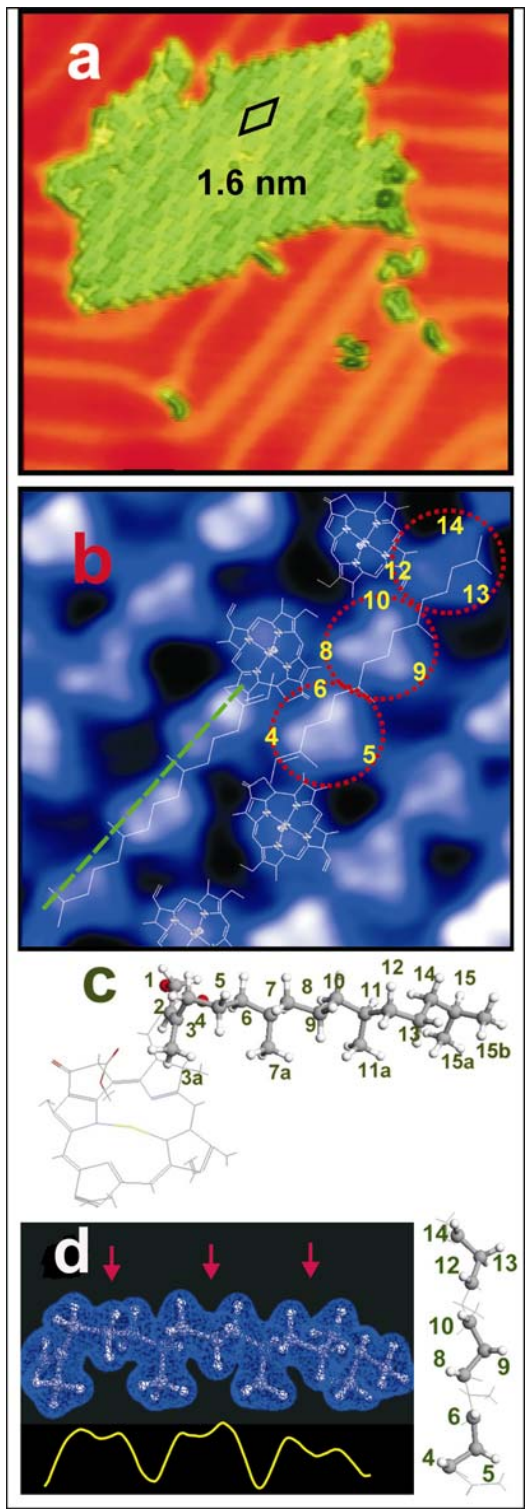

**Fig. 1.** Chlorophyll-a structure: (*a*) A self-assembled chlorophyll-a cluster having a rhombohedral unit cell of 1.6 nm length, and scattered single molecules on Au(111) surface with the Au(111) herringbone reconstruction in the background. [30 x 28 nm$^2$ area, $V_t$ = 1.5 V, $I_t$ = 280 pA]. (*b*) Chlorophyll-a molecules inside the cluster show three bright triangular units (shown with red circle) contributed from the phytyl-chain. The carbon atoms of these CH groups are labelled. [2.8 x 2.5 nm$^2$ area, $V_t$ = 0.3 V, $I_t$ = 280 pA]. (*c*) The calculated geometry of chlorophyll-a using the parametric method shows the porphyrin 'head' and phytyl 'tail' positions. (*d*) A side view of the phytyl charge-density plot. The red arrows indicate the locations of three triangular units seen in (*b*). A yellow STM line scan (1.6 nm length) taken along the green line in (*b*) reveals similar profile as the calculated density. (Right) The top view of the phytyl chain structure shows the elevated CH groups seen in (*b*).

The calculated structure of a free standing chlorophyll-a using the parametric (PM3) method (15) shows that the phytyl chain is folded on the porphyrin unit (Fig. 1*c*). Chlorophyll-a adsorbs on Au(111) by maintaining similar structure: The porphyrin unit is lying flat on the surface while the phytyl chain is folded on top and elevated from the surface by the support of four $CH_3$ groups. In the high resolution STM images, the phytyl appears as a chain of three bright triangular units, which are assigned as the elevated C-H groups corresponding to the following carbon atoms: (C4, C5, C6), (C8, C9, C10), and (C12, C13, C14), respectively (Fig. 1*c*, and 1*d*). The carbon atoms C3, C7, C11, and C15 are attached to the $CH_3$ groups, and are located slightly lower than their neighbors in this conformation (Fig. 1*d*).

Isolated chlorophyll-a molecules on Au(111) show similar structure as in the clusters (Fig. 2*a*) with less detailed features: Instead of three triangular units, it appears as three lobes in STM images. This might be partly due to a higher mobility of the phytyl in isolated molecules during STM scanning as oppose to the ones in the clusters, which are locked-in by neighboring molecules. In this adsorption geometry, the binding of phytyl chain to the surface is weak, and thus it is relatively easy to change its conformation by using inelastic tunneling spectroscopy (IETS) scheme (2).

Four conformations of chlorophyll-a, marked as '1, 2, 3', and '4' (Fig. 2*a* to 2*d*), have been selectively switched by injecting tunneling electrons from the STM-tip. Each switching step involves 60º bending of the phytyl-chain by positioning the STM-tip at one of the two locations indicated as "A" and "B" in Fig. 2. For instance, to switch between the conformations '1' and '2' (Fig. 2*a* and 2*b*), the STM-tip is placed at a fixed height above the phytyl-chain near "A", and tunneling electrons are injected into the molecule using 1.5 V bias. Switching the conformation '2' to '3' and '3' to '4' are also realized by using a similar process. The '2-3' switching (Fig. 2*b*, and 2c) involves 60º bending at "B" while the phytyl-chain angle at "A" remains unchanged. The '3-4' switching is done by bending the phytyl at "A" again. When the molecule is in conformation '2', it can be switched either to conformation '1' or '3' by bending of the phytyl at separate locations "A" and "B". Similarly, '3-2', and '3-4' switching processes can be selectively performed by bending the phytyl at separate locations, "B" and "A", respectively.

To understand the structural changes during switching, we have performed geometrically relaxed calculations for several free standing chlorophyll-a structures using parametric (PM3) method (15). The calculations reveal that the three bent molecular conformations observed here are caused by rotations of parts of the phytyl-chain. The '1-2' switching involves bending of phytyl at C12 joint. This is caused by rotation of $\sigma_{C11-C12}$ bond (indicated in Fig.2 with a red arrow in '1'), which rotates the last part of the phytyl (from C12 to C15) into a counter clockwise direction resulting in a 60º tail bending (Fig. 2). The '2-3' switching is induced by bending the phytyl at C8 joint by rotating $\sigma_{C7-C8}$ bond (shown in Fig. 2 with a blue arrow in '2'). This rotates a part of the phytyl, from C8 to C15, into a counter clockwise direction. A clockwise rotation of the end part of phytyl at C13 in conformation '3' switches to the conformation '4' (indicated in Fig. 2 with an orange arrow in '3'). Since the phytyl chain is lifted up from the surface by the support of $CH_3$ groups, such chain rotations can easily take place. The calculations revealed that chlorophyll-a can have a rich variety of conformations. We conclude that our ability to switch only four chlorophyll-a conformations on Au(111) is due to the limit of phytyl tail rotation imposed by the surface. By comparing the calculated results with the experiment, we find the location "A" (Fig. 2) as the C12 and C13 region while "B" as the C8 joint.

During the tip induced switching process, the tunneling current is recorded as a function of time and the conformation changes can be recognized from the abrupt changes in current intensity (Fig. 3a). By increasing the process duration, the molecule can be switched between the two conformations back and forth producing a two-step current signal (Fig. 4b) (16-19), which functions like a toggle-switch. In fact, the conformation switching is occasionally observed at the tunnelling voltages as low as 0.5 V. In general, rotational excitations of a molecule require lower energies than electronic excitations. This low voltage requirement is in agreement with the theoretical observations, which reveal that all the tail bending are caused by rotation of $\sigma_{C-C}$ bonds. As an example, an I-V curve of an isolated chlorophyll-a revealing the current fluctuation above 0.8 V due to the conformation changes (20) is illustrated in Fig. 3c. Here, as the current increases at the larger voltages, the frequency of switching also increases (20). For a fixed bias, the average switching frequency can be adjusted by varying the tunneling current only. The increase or decrease in the current induces faster or slower switching rate, respectively.

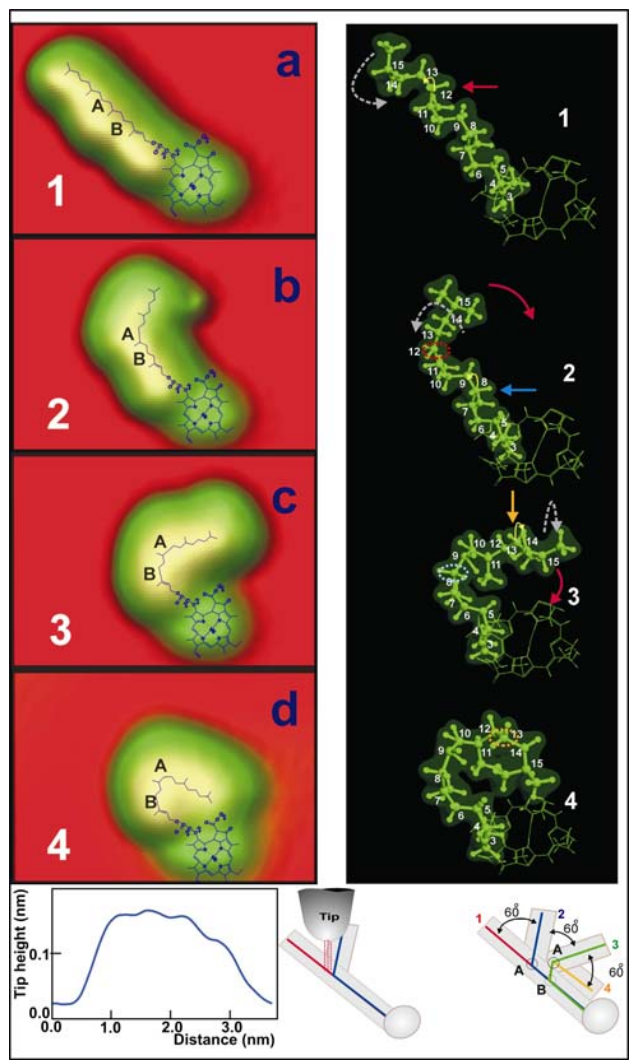

**Fig. 2.** Four-step molecular conformation switching. STM images (left) and calculated structures (right). **(a)** A straight-tail conformation "1" can be switched into a bent-tail conformation "2", **(b),** by changing the tail angle at "A". This is caused by a counter clockwise rotation of phytyl chain at the location shown with a red arrow in the calculated image. Another 60° bending at "B" in STM image (left) changes conformation "2" to "3",**(c)**. This tail-bending is due to a part of phytyl rotation indicated with a blue arrow in the calculated image (right). Further 60° bent at "A" changes the conformation "3" to "4", **(d)**. This is realized by a clockwise rotation of phytyl shown with an orange arrow in the calculated image (right). The cartoon drawings below demonstrate the STM tip-induced switching procedure, and the tail bending locations, "A" and "B".

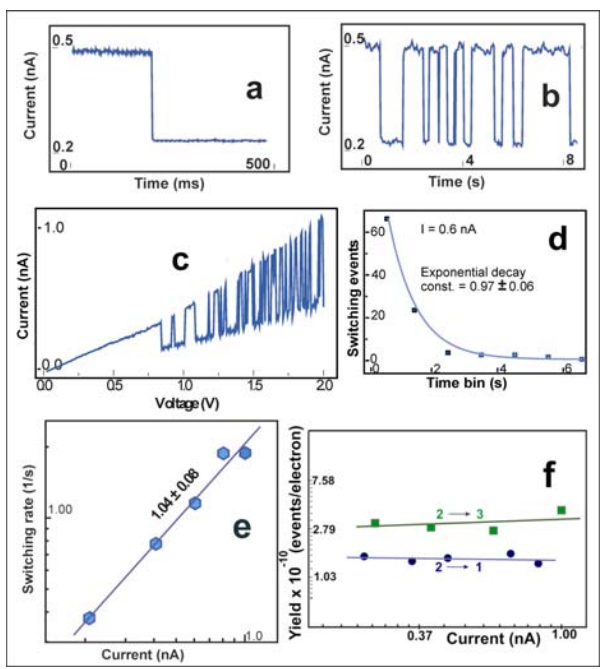

**Fig. 3.** Switching operation and mechanism: **(a),** An abrupt change in the tunneling current is associated with '1-2' switching. **(b),** For a longer electron injection period, a two-step multiple-switching signal is observed. **(c),** An I-V curve of chlorophyll-a shows fluctuation of current above 0.8 V due to switching between '1' and '2'. **(d),** A binned distribution of the switching events corresponding to 0.6 nA. **(e),** A logarithmic switching rate vs. current graph shows a unity slope. **(f),** The yield plots for '2-1' and '2-3' switching events.

To understand the tunnelling electron induced switching process, we have performed statistical analyses taken over 1200 switching events using 1.5 V bias, and at different currents for the four molecule conformations. Fig. 3*d* shows the plot of conformation changes vs. time for '1-2' switching with a fixed current of 0.6 nA. The time constant of this exponential decay has been determined as $0.97 \pm 0.06$ s. The switching rates are determined from the inverse of the time constants. The linear dependence of switching rate on the tunneling current is illustrated in Fig. 3*e*, where each data point is determined by plotting an exponential curve shown in Fig. 3*d*. The rate "R", and tunneling current "I" are related as

$$R \propto I^N,$$

where "N" is the number of inelastic tunneling electrons involved in the energy transfer process (21-23) to the molecule that induces conformation switching. From the slope of this curve, N is determined as $\approx 1$. Therefore, this switching process is initiated by a single tunneling electron energy transfer process. Similar results have been found for the '2-3' and '3-4' switching processes, and thus single tunneling electron energy transfer causes all the switching events described in this report. We have also determined quantum yields for the switching events. Fig. 3*f* provides the yield plots for '2-1' and '2-3' switching events. Here, the yield "y" is related to the switching rate "R" and current "I" as,

$$y = Re/I$$

where "e" is the electronic charge. The determined yields are $3.2 \times 10^{-10}$ for switching from '2' to '3', and $1.5 \times 10^{-10}$ for the switching from '2' to '1', respectively. Almost zero slopes of the yield plots further verify the single electron energy transfer process (24). The measured yields for '3-2' and '3-4' switching events are $3.8 \times 10^{-10}$, and $3.3 \times 10^{-10}$, respectively. To demonstrate our ability of control further, we present snapshot images from a STM movie (supporting information) showing the switching of chlorophyll-a conformation between the conformations '1, 2, 3', and '4' in Fig. 4.

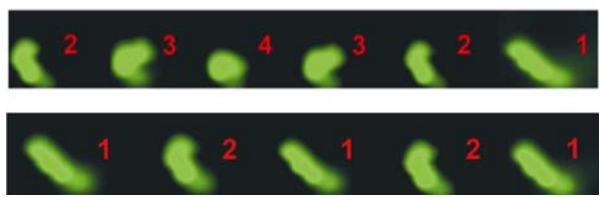

**Fig. 4.** Frames of STM images from a STM movie showing reversible switching of conformations '1,2,3, and 4'.

In summary, we report the first extensive study of single chlorophyll-a molecules and provide detailed information about the structural properties of isolated molecules and self-assembled molecular structures at the atomic level. Our four-step single molecule switching scheme demonstrates that chlorophyll-a may be useful in nanoscale biomechanical devices. Furthermore, this achievement opens a novel route to investigate, and even to control, the conformation changes of biological molecules including some proteins with a sub-molecular resolution using STM manipulation schemes.

We thank Jessica Benson for initiation of the project. We gratefully acknowledge the Ohio University Bionanotechnology Initiative and funding provided by the United States Department of Energy, BES grant number DE-FG02-02ER46012.